\documentclass[12pt]{iopart}
\usepackage[latin1]{inputenc}
\usepackage{epsfig}
\usepackage{multicol}
\usepackage{latexsym,amsfonts}
\def\be{\begin{equation}}
\def\ee{\end{equation}}
\newcommand{\beqa}{\begin{eqnarray}}
\newcommand{\eeqa}{\end{eqnarray}}
\newcommand{\om}{\Omega_M}
\newcommand{\ode}{\Omega_{DE}}
\newcommand{\omo}{\Omega_M^{(0)}}
\newcommand{\odeo}{\Omega_{DE}^{(0)}}
\newcommand{\w}{\omega}
\newcommand{\dl}{\delta}

\begin{document}

\title{ Dark energy parameterizations and their effect on dark halos }

\author{Lamartine Liberato and Rogerio Rosenfeld}

\address{Instituto de F\'{\i}sica Te\'orica- State University of S\~ao
Paulo\\
Rua Pamplona 145, 01405-900, S\~ao Paulo, SP, Brazil.}
\eads{\mailto{liberato@ift.unesp.br}, \mailto{rosenfel@ift.unesp.br}}


\begin{abstract}
There is a plethora of dark energy parameterizations that can fit
current supernovae Ia data. However, this data is only sensitive
to redshifts up to order one. In fact, many of these
parameterizations break down at higher redshifts. In this paper we
study the effect of dark energy models on the formation of dark
halos. We select a couple of dark energy parameterizations which
are sensible at high redshifts and compute their effect on the
evolution of density perturbations in the linear and non-linear
regimes. Using the Press-Schechter formalism we show that they
produce distinguishable signatures in the number counts of dark
halos. Therefore, future observations of galaxy clusters can
provide complementary constraints on the behavior of dark energy.
\end{abstract}

\noindent{\it Keywords\/}: dark energy, dark halos, number counts

\maketitle

\section{Introduction}\label{sec.intro}
$ $

There are strong and converging evidences suggesting that the
universe is nearly flat and that the dominant component of the
energy density today has an unknown nature. Analysis of the
luminosity distance of high redshift Type Ia supernovae has led to
the conclusion that the expansion of the universe is accelerating,
with indications that this acceleration is recent
\cite{riess98,perlmutter98,knop03,riess04,astier05} and are
corroborated by cosmic microwave background radiation
\cite{spergel03cb,wmap3y} and large scale structure observations
\cite{tegmark03a,cole05}.

This suggests that the dominant contribution to the present-day
energy budget can be described by a fluid with equation of state
$\w<-1/3$, usually called ``dark energy''. We do not know for sure
what this dark energy really is and in our ignorance many
parameterizations for the time or redshift dependence of the
equation of state $\w(z)$ have been introduced in order to
observationally constrain its evolution.

Type Ia supernovae data are sensitive to $\w(z)$ only for a small
range of redshifts, typically up to $z = {\cal O}(1)$. This is the
reason for the large allowed regions of parameters related to the
variations of $\w(z)$. In fact, the simplest possibility, a
cosmological constant with $\w=-1$ still provides a good fit to
the data. Therefore, it would be highly desirable to have another
observable, sensitive to higher redshifts, that could break the
degeneracy among the several different proposed parameterizations.

Dark energy has a dramatic effect on the dynamics of the universe,
altering the way in which cosmological structures grow
\cite{percival05}. This offers the possibility that observations
of structure formation may provide a sensitive probe of dark
energy properties that is complementary to information derived
from supernovae data.

The build-up of structure in universes with dark energy is subject
to significant ongoing study
\cite{lahav91,viana95,coble97,wang98,linder03a,mainini03,mota04,
bidgoli05,horellou05,nunes05,bdw05,manera05,delliou06},
particularly through the spherical collapse model \cite{GunnGott}
in association with the Press-Schechter formalism for the mass
function \cite{ps}. In this paper we use these techniques to
investigate cosmologies in which the dark energy component remains
homogeneous on the scales of the structures being modelled and
their consequences to structure formation, in special to the
cluster number counts. In particular, we show that cluster number
counts can be sensitive to the different parameterizations of the
dark energy equation of state.

This paper is organized as follows. In section 2 we show the
parameterizations we will be using to illustrate our results. In
section 3 we study how the linear growth of perturbations is
affected by these models of dark energy. Section 4 is devoted to
the non-linear growth of perturbations in the spherical collapse
model and the computation of the threshold density contrast for
collapse in the different models studied. In section 5 we present
our results for the number counts of dark halos in different mass
bins and also an integrated count in the different models studied.
Finally, our conclusions are presented in section 6.

\section{ Parameterizations for the dark energy equation-of-state }\label{sec:blt}
$ $

The expansion history of the universe is determined by the Hubble
parameter, $H(t)=\dot a/a$, where $a(t)$ is the scale factor
($a(t_0)\equiv a_0 = 1$ today). We assume that the dark energy has
an equation of state relating its pressure $p_{DE}$ and density
$\rho_{DE}$ at a particular instant determined by $a(t)$ given by
$p_{DE}=\w (a)\rho_{DE}$ and that matter is pressureless,
$p_{M}=0$. For general $\w (a)$, the expansion rate of the
universe is governed by the Friedman equation \be
   \frac{H^2(a)}{H_0^2}
   = \omo \, a^{-3}+ \Omega_K \, a^{-2}+\odeo \; e^{f(a)}\,,
   \label{h2}
\ee where $\omo$, $\Omega_K$ and $\odeo$ are the current density
parameters due to non-relativistic matter (baryonic and
non-baryonic), curvature and dark energy. In the following we
assume a flat universe, $\Omega_K\equiv(1-\om-\ode)=0$. $H_0$ is
the Hubble constant and the function $f(a)$ is determined by the
dark energy equation of state:
 \be
   f(a) = 3\int_a^1[1+\w (a')]d\ln a'\,.
   \label{f}
\ee

The matter density $\om(a)$ and dark energy density $\ode(a)$ are
functions of $a$:
 \be
   \om(a)=\omo \, a^{-3} \, \frac{H_0^2}{H^2(a)}\,,
   \label{omegam}
\ee
and
\be
   \ode (a)=\odeo \; e^{f(a)} \; \frac{H_0^2}{H^2(a)}\,.
   \label{omegade}
\ee
Throughout this paper we will adopt $\omo = 0.25$, $\odeo =
0.75$, $H_0 = 72$ km s$^{-1}$ Mpc$^{-1}$ \cite{wmap3y} and use
either the scale parameter $a$ or the corresponding redshift $z =
(1-a)/a$ to describe the evolution of the different parameters.

Recently discovered Type Ia supernovae (SNeIa)
\cite{riess04,astier05} provide conclusive evidence of the
decelerating universe in the past $(z>0.5)$ evolving into the
present day accelerating universe. Thus the existence of dark
energy, which accelerates the cosmic expansion, has been firmly
established and the magnitude of its energy density today has been
accurately measured. The goal is now to determine the behavior of
the dark energy density and its equation of state at different
cosmic epochs. The simplest and most natural candidate for dark
energy is the cosmological constant $\Lambda$ with a constant
energy density $\Omega_{\Lambda}$ and a fixed equation of state
parameter $\w = -1$. It still provides a good fit to SNeIa data.

The exact functional form of the equation of state $\w (z)$ should
follow from a fundamental theory for the dark energy. The best
studied case is the so-called quintessence models, where dark
energy results from a scalar field rolling down a potential. In
this class of models, the equation of state is strictly in the
region $-1 < \w <1$. However, more complicated models with
multiple scalar fields or non-canonical kinetic energy can have
$\w<-1$.

In the absence of a physically well-motivated fundamental theory
for dark energy, it has become common practice to adopt parametric
forms of $\w(z)$ and to use SNeIa data to find the allowed regions
in the parameter space for different parameterizations
\cite{linder03a,huterer00,
jassal04,chevallier00,weller01,sahni03,lazkoz05}.

Many parameterizations were proposed to fit the observations and
some of them are shown in Table~\ref{models}.

\begin{table*}[h]
\begin{minipage}{160mm}
\begin{center}
\caption{\sf Different parameterizations for dark energy (some of
these models \\are from \cite{lazkoz05}) }\label{models}
\vspace{-10pt}
\begin{tabular}{cll}
                           \\ \hline \hline
{\bf Model}  & $\bf{ H^2(z)}$ or $\bf{\w (z)}$ & {\bf Parameters }
                           \\ \hline
{\bf Ia}     & $\w (z) = \w_0 + \w_1\frac{z}{(1+z)²} = \w_0 + \w_1 (1-a)a$
             & $\w_0= -1.3$  \\
             &&$\w_1= 4$  \\ 

{\bf Ib}     & $\w (z) = \w_0 + \w_1\frac{z}{(1+z)²} = \w_0 + \w_1 (1-a)a$
             & $\w_0= -1.3$   \\
             &&$\w_1= -2$  \\ 

{\bf II}     & $\w (z) = \w_0 + \w_1\frac{z}{(1+z)^{1.8}}
                       = \w_0 + \w_1 (1-a)a^{0.8}$
             & $\w_0= -1.3$    \\
             &&$\w_1= 4$     \\ 

{\bf III}    & $H^2 (z)=H_0^2 \ [\omo (1+z)^3 + \odeo +$
             & $a_1=0.13$      \\
             & $+ a_1 (1+z)^3[\cos(a_2 z+a_3\pi)-\cos(a_3\pi)]]$
             & $a_2=6.83$      \\
             &&$a_3=4.57$      \\ 

{\bf IV}     & $H^2 (z)=H_0^2 \{\omo (1+z)^3 + a_1(1+z)+$
             & $a_1=-4.16$     \\
             & $a_2(1+z)^2+( \odeo -a_1-a_2)\} $
             & $a_2=1.67$      \\

{\bf V}      & $H^2 (z)=H_0^2 [\omo (1+z)^3 -$
               $\sqrt{a_1 + a_2 (1 + z)^3}$
             & $a_1=29.08$     \\
             & $( \odeo +\sqrt{a_1 + a_2})]$
             & $a_2=-0.097$    \\

{\bf VI}     & $\w (z) = \frac{\w_0}{1+b \; ln (1+z)}$
             & $\w_0= -1$    \\
             &&$b= 0.25$        \\

{\bf VII}    & $\w (z) = \w_0 + \w_1 \; z = \w_0 + \w_1 \frac{1-a}{a}$
             & $\w_0= -1.4$  \\
             &&$\w_1= 1.67$  \\ 

{\bf VIII}   & $\w (z) = \w_0 + \w_1\frac{z}{1+z}
                       = \w_0 + \w_1 (1-a)$
             & $\w_0= -1.6$  \\
             &&$\w_1= 3.3$  \\ 

{\bf $\Lambda$ } & $\w = -1$ $\;\;\;$ or  $\;\;\;$
                   $H^2 (z)=H_0^2 [\omo (1+z)^3 + \odeo]$
                 & -  \\ \hline\hline
\end{tabular}
\end{center}
\end{minipage}
\end{table*}

\begin{figure}[h]
   \centerline{\psfig{file=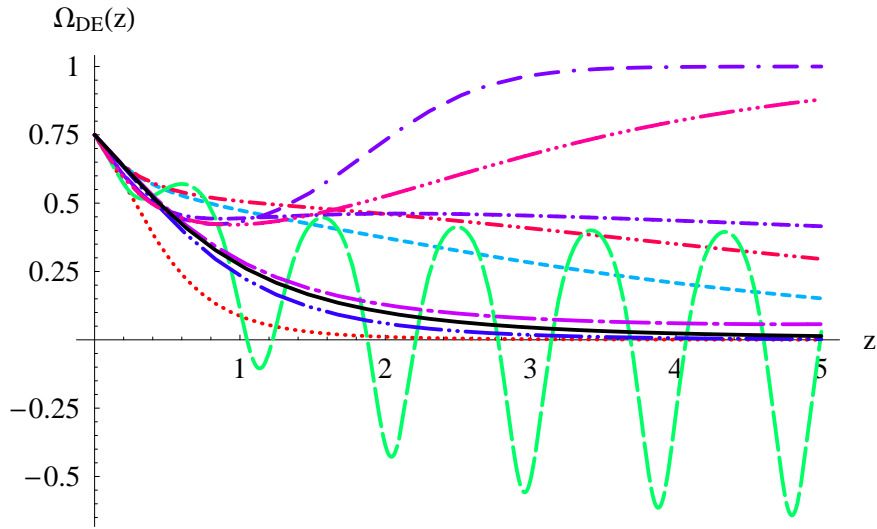,scale=0.85,angle=0}}
   \caption{\sf Evolution of the dark energy $\ode$ with redshift for
   parameterizations of Table~1. The lines are as follows: $\Lambda$CDM
   (solid), Ia (short dashed), Ib (dotted), II (double dot dashed ),
    III (long dashed), IV (dot short dashed), V (dot long dashed),
    VI (double dot long dashed), VII (double dashed doted), VIII
    (triple dot dashed).
   }\label{f.omegade}
\end{figure}

\begin{figure}[h]
   \centerline{\psfig{file=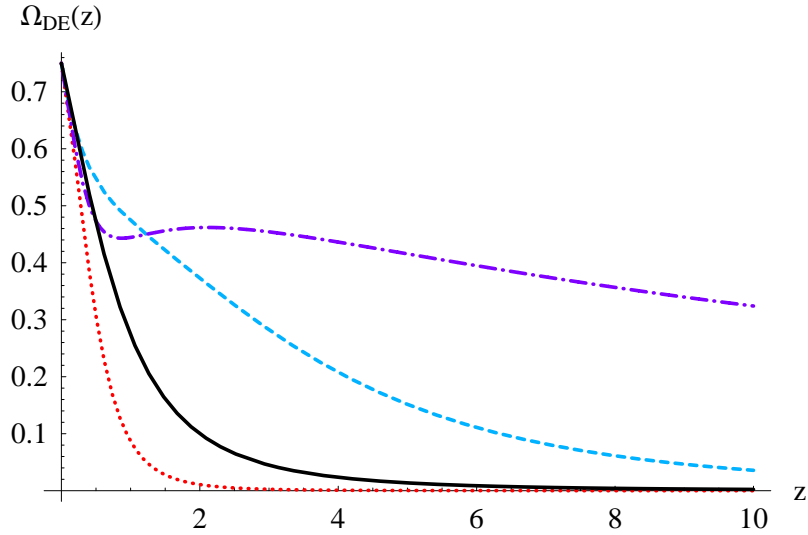,scale=0.85,angle=0}}
   \caption{\sf Evolution of the dark energy $\ode$ with redshift for
   selected models of Table~1. The lines are as follows: $\Lambda$CDM
   (solid), Ia (short dashed), Ib (dotted) and
   IV (dot short dashed).}\label{f.omegade2}
\end{figure}

\begin{figure}[h]
   \centerline{\psfig{file=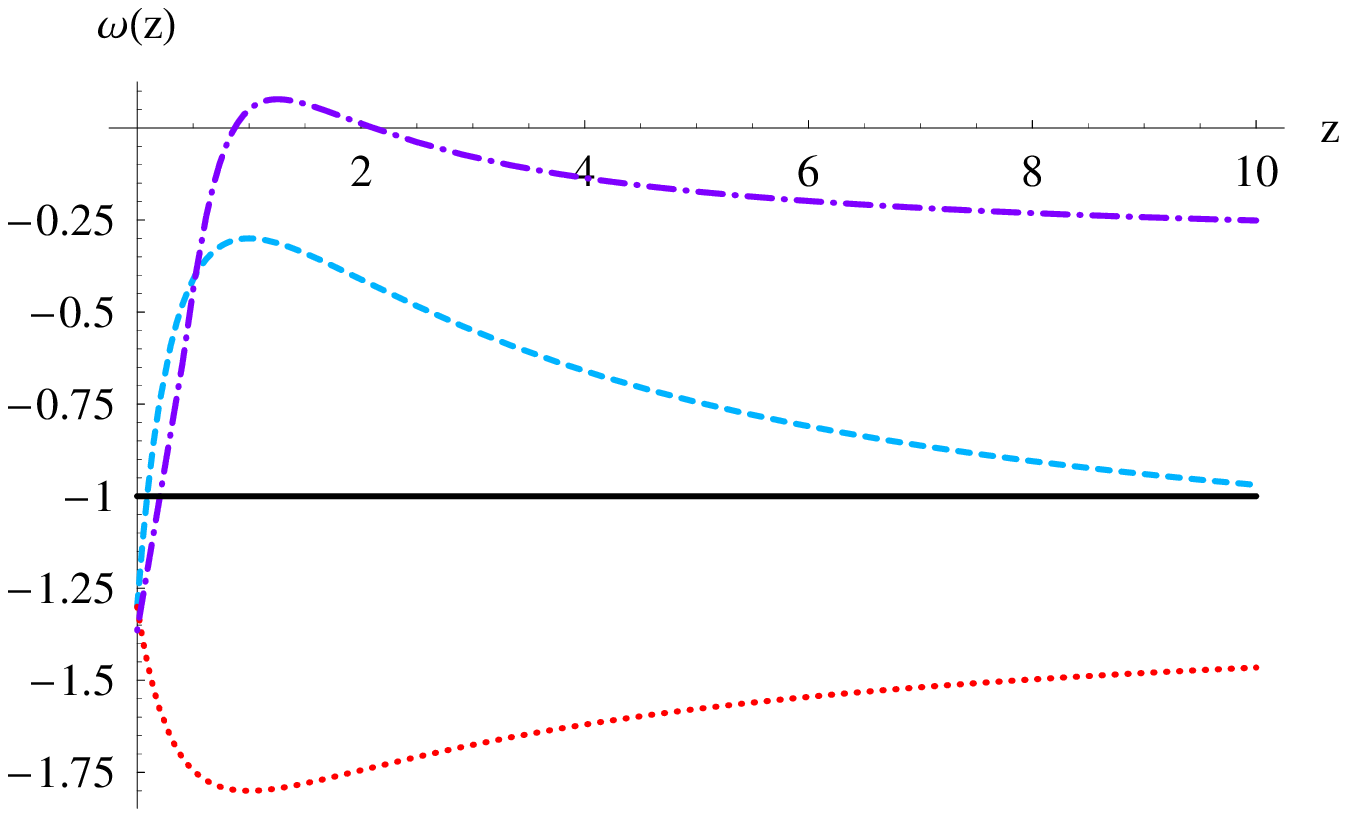,scale=0.85,angle=0}}
   \caption{\sf Equation of state evolution $\w(z)$ with redshift for
   selected models of Table~1. Lines are the same as in figure \ref{f.omegade2}.}\label{f.w}
\end{figure}


The parameterizations or models shown in Table~\ref{models} are
obtained from several sources. Models Ia and Ib are the same
parameterization with the same central value of $w_0$ but with
values for $w_1$ at the extrema of the range allowed by SNLS and
WMAP \cite{jassal06}. Model II is a slight modification
of model I. Models III, IV and V are an oscillating, a quadratic
polynomial \cite{sahni03} and a brane-motivated ansatze for $H(z)$, respectively
and we used the central values of the parameters \cite{lazkoz05}.
Model VI was proposed by Wetterich \cite{wetterich04} and its best
fit parameters are taken from \cite{bidgoli05}. The
parameterizations VII \cite{huterer00,weller01} and VIII
\cite{linder03a,chevallier00} are first order Taylor expansions
around $z=0$ and around $a=1$, respectively. In figure
\ref{f.omegade} we show the behavior of the dark energy densities
resulting from these parameterizations up to $z=5$ and we see that
most of them present problems such as a too large dark energy
component for redshifts higher than $z=1$.

For our study we select only a couple of models with decreasing
contribution of $\ode(z)$ in the past. In order to contrast with
the $\Lambda$CDM model, we choose models with energy densities
both above and below the one given by a cosmological constant.
These are models Ia, Ib and IV, whose energy densities and
equations of state are depicted in figure \ref{f.omegade2} and
figure \ref{f.w}.

In the next sections we investigate the consequences of these
different models to cosmological parameters that are relevant for
large scale structure formation in the universe.

\section{ Linear Perturbation Theory }\label{s.lpt}
$ $

We assume that the dark energy component is smooth on scales
smaller than the horizon \cite{ma99}. In this case we only need to
consider perturbations to the non-relativistic matter component.
Dark energy only alters the background evolution and the equation
of perturbed density contrast (for perturbations larger than the
Jeans length) is \cite{pad93}
\be
   \ddot\dl +2H(t)\,\dot\dl -{3\over 2}H(t)^2\om(t) \,\dl = 0\,,
   \label{deltat}
\ee
where $\dl$ is the fractional matter density perturbation.

The growth function is defined as the ratio of the perturbation
amplitude at some scale factor relative to some fixed scale
factor, $D=\dl(a)/\dl(a_0)$, and its evolution equation can be
written as
 \be
    D'' + {3\over2}\left[1-{\w (a)\over 1+X(a)}\right]\,{D'\over a}-{3\over2}
     \,{X(a)\over 1+X(a)}\,{D\over a^2}=0\,,
     \label{d}
\ee where the prime denotes derivative with respect to the scale
factor $a$ and $X(a)$ is the ratio of the matter density to the
dark energy density:
 \be
   X(a) = {\omo\over 1-\omo} e^{-3\int_a^1 d\ln a'\,\w (a')}\,.
   \label{x}
\ee

\begin{figure}[thb]
   \centerline{\psfig{file=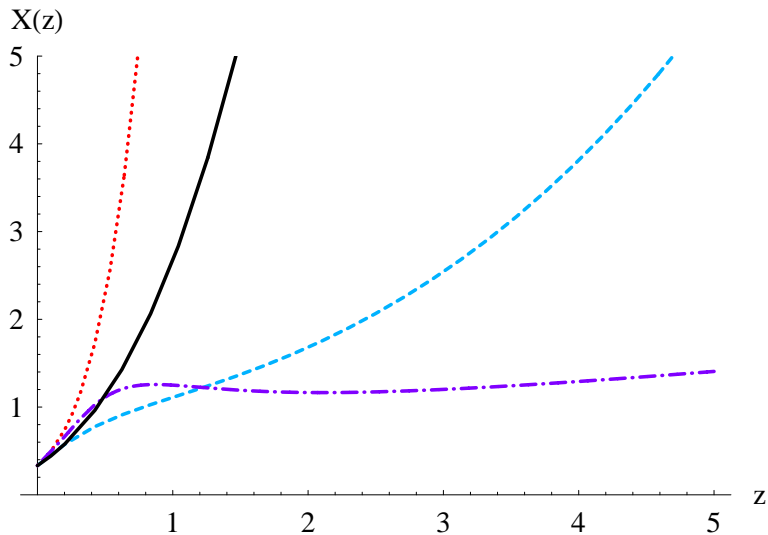,scale=0.85,angle=0}}
   \caption{\sf Behavior of the $X$ with redshift for dark energy scenarios
   considered. Lines are the same as in figure \ref{f.omegade2}.}
   \label{f.xb}
\end{figure}

For large $X$ one recovers the matter dominated behavior $D\sim
a$. The ratio of the matter density to the dark energy density
$X(z)$ can be seen in figure \ref{f.xb}. Notice that for all
models except model IV the dark energy contribution decreases
rapidly with redshift. In model IV the dark energy density
decreases slowly but for redshifts larger than $0.5$ this
parameterization does not accelerate the universe.

\begin{figure}[h]
   \centerline{\psfig{file=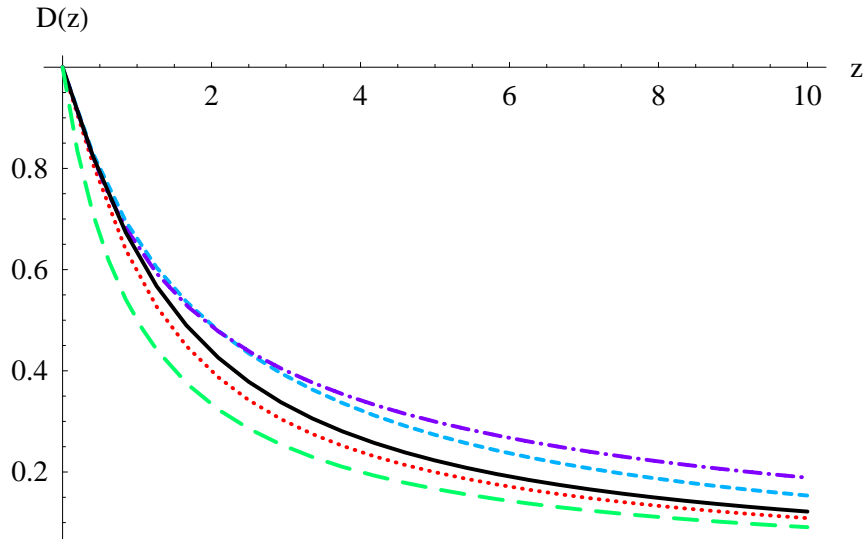,scale=0.85,angle=0}}
   \caption{\sf Evolution of the growth factor $D(z)$ with redshift for
   selected dark energy models(lines are the same as in figure (\ref{f.omegade2})) and for pure dark matter case (in which case
   $D(z) = a(z) = \frac{1}{1+z}$) (long dashed line).}\label{f.dz}
\end{figure}

In figure \ref{f.dz} we show the evolution of the growth function
$D(z)$ for the selected dark energy models and for pure dark
matter case. Notice that, as expected, larger perturbations in the
past are needed to arise at the same amplitude today for models
where dark energy is more important since negative pressure tends
to inhibit the growth of perturbations.

\section{ Non-Linear Evolution of Perturbations }\label{s.nl}
$ $

\begin{figure}[h]
   \centerline{\psfig{file=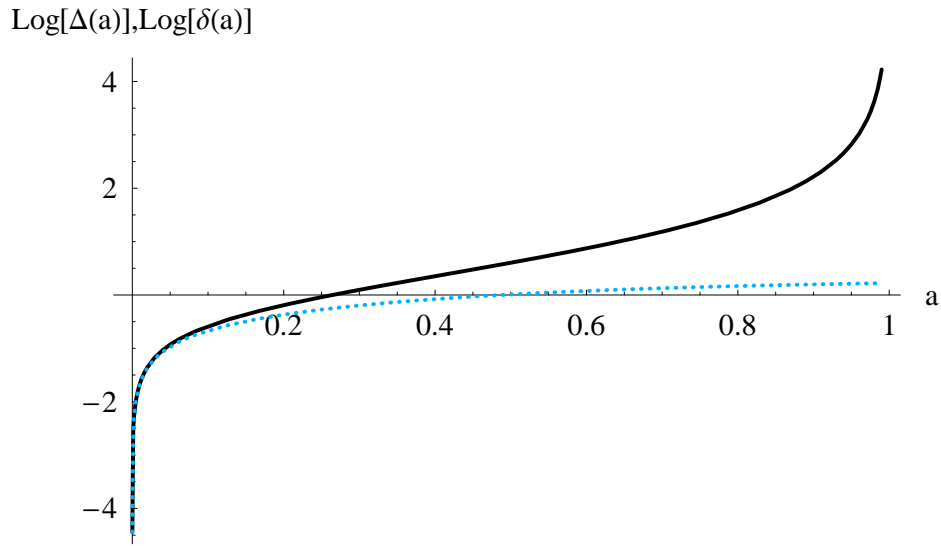,scale=0.95,angle=0}}
   \caption{\sf Evolution of the linear (dotted line) and
   non-linear (solid line) density contrasts in $\Lambda$CDM
   for an initial value chosen so that the collapse occurs today.
}\label{nonlinear}
\end{figure}

In order to describe the non-linear evolution of the density
perturbations we adopt the spherical collapse model
\cite{GunnGott} where the radius $R(t)$ of a spherical homogeneous
overdensity region obeys the curvature-independent Raychaudhuri
equation:
\be \ddot{R} = -\frac{3}{2} \odeo (w(a) + 1/3) e^{f(a)} R -
\frac{1}{2} \omo (1+ \Delta_i) \frac{1}{R^2}, \ee where time is
measured in units of $1/H_0$ and $\Delta_i$ is the initial
overdensity in the sphere. We numerically solved this equation for
an initial time $t_i$ where $a(t_i) = 10^{-5}$, with initial
conditions chosen so that the sphere is initially in the Hubble
flow, $R(t_i) = a(t_i)$, $\dot{R}(t_i) = \dot{a}(t_i)$.  We find
the values of $\Delta_i$ such that the collapse occurs today. The
linear evolution of $\delta_i = \Delta_i$ until today results in
the critical linear density contrast parameter $\delta_c$ which is
important in the Press-Schechter \cite{ps} formalism discussed
below. In figure \ref{nonlinear} we show an example in the
$\Lambda$CDM model of the linear and non-linear evolution for an
initial overdensity of $\Delta_i = 10^{-4.446}$, chosen so that
the collapse occurs today.

In an Einstein-de Sitter universe, an exact value of $\delta_c =
1.686$ is obtained \cite{GunnGott} and we verified that this value
is fairly independent of the background cosmology
\cite{mainini03}.

\begin{figure}[h]
   \centerline{\psfig{file=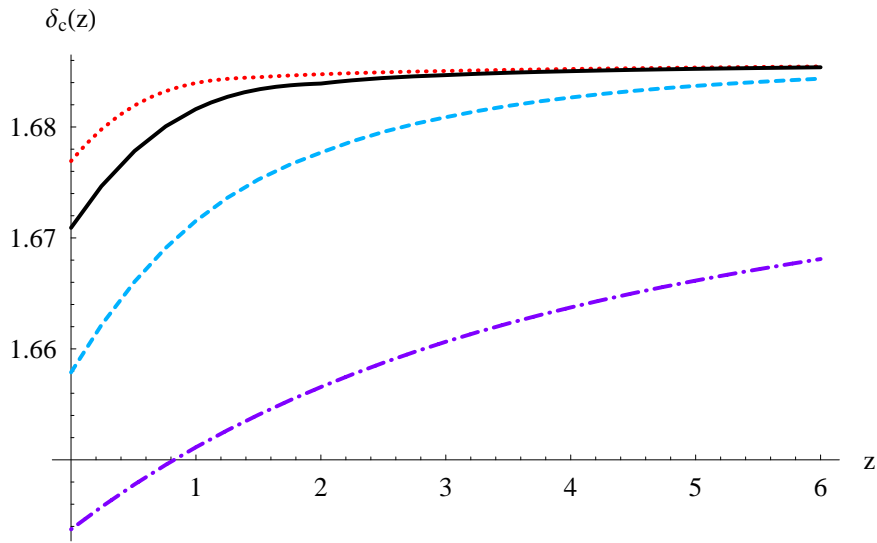,scale=0.85,angle=0}}
   \caption{\sf Threshold overdensity for different redshifts of
   collapse for the models considered. Lines are the same as in figure \ref{f.omegade2}.}\label{dc}
\end{figure}

We have also computed the values of $\delta_c(z)$ for different
collapse redshifts. Our results, shown in figure \ref{dc}, find no
large numerical differences among the models, in agreement with
\cite{percival05,NunesMota}. We use these values of $\delta_c(z)$
to compute the halo abundances in the next section.

\section{Mass Function and Cluster Number Counts}\label{s.mf}
$ $

The most reliable method to study the cluster abundance in the
universe is through numerical simulations. However, there is an
analytical approximation, the Press-Schechter formalism \cite{ps},
that has been shown to fairly reproduce the results of N-body
simulations \cite{LaceyCole}. There are more recent and better
approximations with extra free parameters \cite{Jenkins} but as an
initial step we will use the well-known Press-Schechter approach.

The basic premisses of the Press-Schechter formalism is to assume
that the fraction of mass in the universe contained in
gravitationally bound systems with masses greater than $M$ is
given by the fraction of space where the linearly evolved density
contrast exceeds a threshold $\delta_c$, defined in the previous
section, and that the density contrast is normally distributed
with zero mean and variance $\sigma^2(M)$, the root-mean-squared
value of the density contrast $\delta$ at scales containing a mass
$M$. Therefore, it is assumed that for a massive sphere to undergo
gravitational collapse at a redshift $z$ its linear overdensity
should exceed a threshold $\delta_c(z)$. Notice that only linear
quantities are used in this formalism.

These assumptions lead to the well-known analytical formula for
the comoving number density of collapsed halos of mass in the
range $M$ and $M+ dM$ at a given redshift $z$: \be
   \frac{dn}{dM} = -\sqrt{\frac{2}{\pi}}\frac{\rho_{\rm m0}}{M}
   \frac{\dl_c(z)}{\sigma (M,z)} \frac{d\ln\sigma (M,z)}{d M}
   \exp\left[-\frac{\dl_c^2(z)}{2\sigma^2 (M,z)}\right] \,,
   \label{mf}
\ee where $\rho_{\rm m0}$ is the present matter mean density of
the universe and $\delta_c(z)$ is the linearly extrapolated
density threshold above which structures collapse, {\it i.e.},
$\delta_c(z) = \dl (z = z_{\rm col})$.

The quantity
\be
   \sigma (M,z) = D(z)\sigma_M
   \label{sigmamz}
\ee is the linear theory {\it rms} density fluctuation in spheres
of comoving radius $R$ containing the mass $M$. The smoothing
scale $R$ is often specified by the mass within the volume defined
by the window function at the present time, see {\it e.g.}
\cite{peebles80}. In our analysis we use the fit given by
\cite{viana95}
\be
   \sigma_M = \sigma_8 \left(\frac{M}{M_8}\right)^{-\gamma(M)/3} \,,
   \label{sigmam}
\ee where $M_8 = 6 \times 10^{14} \omo h^{-1} M_{\odot}$ is the
mass inside a sphere of radius $R_8 = 8 h^{-1} {\rm Mpc}$, and
$\sigma_8$ is the variance of the over-density field smoothed on a
scale of size $R_8$. The index $\gamma$ is a function of the mass
scale and the shape parameter, $\Gamma = \omo h\; e^{-\Omega_b -
\Omega_b/\omo}$ ($\Omega_b = 0.05$ is the baryonic density parameter), of
the matter power spectrum \cite{viana95} \be
   \gamma (M) = (0.3 \Gamma + 0.2) \left[ 2.92 + \frac{1}{3} \log
   \left(\frac{M}{M_8}\right) \right] \,.
   \label{gamma}
\ee

\begin{figure}[h]
   \centerline{\psfig{file=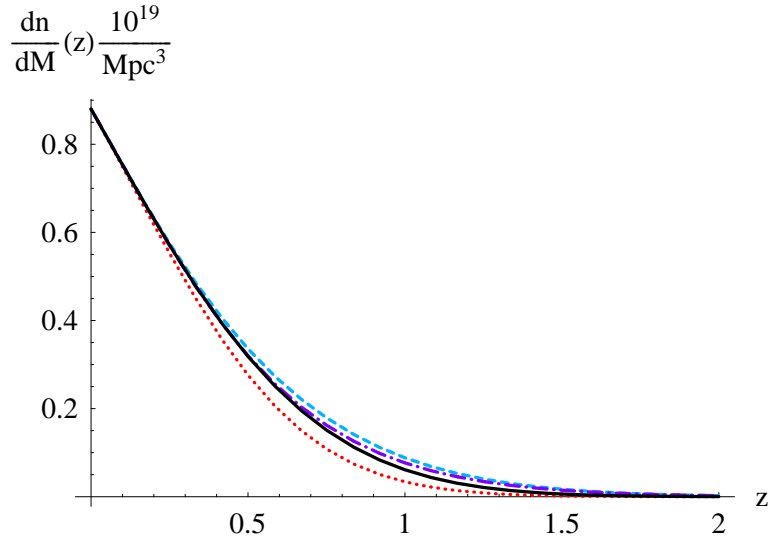,scale=0.85,angle=0}}
   \caption{\sf Press-Schechter mass functions for the different
   models with the $\sigma_8$ normalization. Lines are the same as in figure \ref{f.omegade2}.}\label{massfunction}
\end{figure}

Denoting $\tilde{\gamma} (M) = \frac {d\ln\sigma (M,z)}{d M}$
(notice that it is $z$ independent),
 \be
   \tilde{\gamma} (M) = (0.3 \Gamma + 0.2) \left[ 2.92 + \frac{2}{3} \log
   \left(\frac{M}{M_8}\right) \right] \,,
   \label{gammatilde}
\ee
we can rewrite (\ref{mf}) as
\be
   \frac{dn}{dM} = -\sqrt{\frac{2}{\pi}}\frac{\rho_{\rm m 0}}{M}
   \frac{\dl_c(z)}{\sigma (M,z)} \,\tilde{\gamma} (M)\,
   \exp\left[-\frac{\dl_c(z)^2}{2\sigma (M,z)^2}\right] \,.
   \label{mf2}
\ee In our study we use $\Gamma = 0.144$ \cite{wmap3y}. For a
fixed $\sigma_8$ (power spectrum normalization) the predicted
number density of dark matter halos given by the above formula is
uniquely affected by the dark energy models through the ratio
$\dl_c(z)/D(z)$. In order to compare the different models, we will
normalize to mass function to the same value today, that is, we
will require \be \sigma_{8,M} =
    \frac{\delta_{c,M}(z=0)}{\delta_{c,\Lambda}(z=0)} \sigma_{8,\Lambda}\,,
\ee where the label $M$ indicates a given model and we use
$\sigma_{8,\Lambda} = 0.76$ \cite{wmap3y}. We show in figure
\ref{massfunction} the resulting mass functions for the different
models.

\begin{figure}[h]
   \centerline{\psfig{file=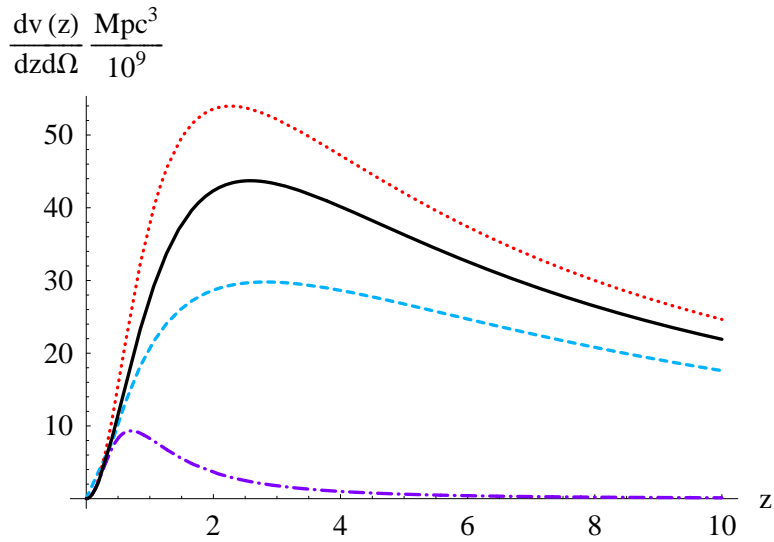,scale=0.85,angle=0}}
   \caption{\sf Evolution of the comoving volume element with redshift
   for the four dark energy scenarios
   considered in this paper. Lines are the same as in figure \ref{f.omegade2}.}
   \label{f.v}
\end{figure}

The effect of dark energy on the number of dark matter halos is
studied by computing two quantities. The first is the all sky
number of halos per unit of redshift, in a given mass bin \be
   {\cal N}^{\rm bin}\equiv\frac{dN}{dz} =
   \int_{4\pi} d\Omega\int_{M_{\rm inf}}^{M_{\rm sup}}
   \frac{d n}{d M} ~\frac{d V}{d z d\Omega} ~dM \,,
   \label{nbin}
\ee
where the comoving volume element is given by
\be
   dV/dz d\Omega = r^2(z)/H(z),
\ee where $r(z) = \int_0^z H^{-1}(x) dx$ is the comoving distance.
The redshift evolution of the comoving volume element $dV/dz
d\Omega $ for different models of dark energy is shown in figure
\ref{f.v}. Note that the comoving volume element does not depend
on the growth factor of the perturbation $D(z)$, but only on the
cosmological background. The comoving volume element is larger for
more negative equation-of-state, since this implies larger
acceleration, see figure \ref{f.w}.
\begin{figure}[h]
   \centerline{\psfig{file=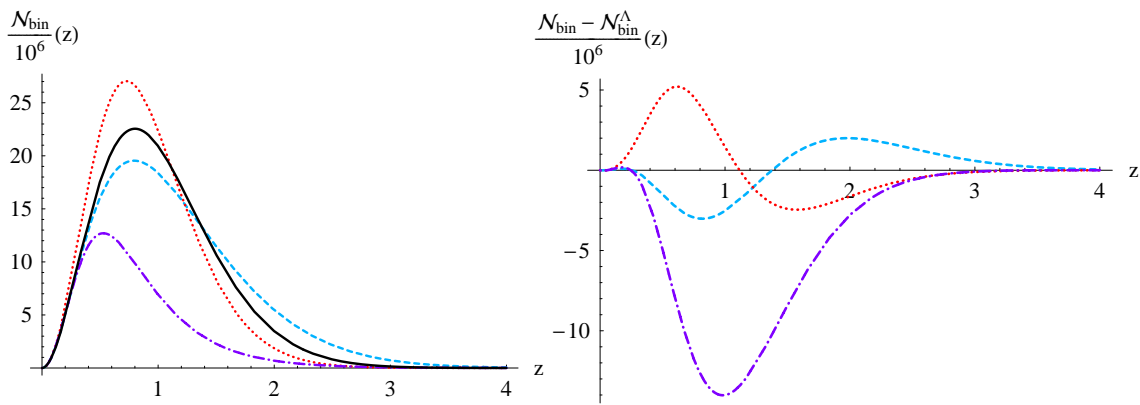,scale=0.65,angle=0}}
   \caption{\sf Evolution of number counts in mass bins with redshift
   for objects with mass within the range
   $10^{13}<M/(h^{-1}M_\odot )<10^{14}$. Notice the normalization factor of
   $\frac{\cal N}{10^6}$. Lines are the same as in figure \ref{f.omegade2}.}
   \label{f.nbin13-14}
\end{figure}
\begin{figure}[h]
   \centerline{\psfig{file=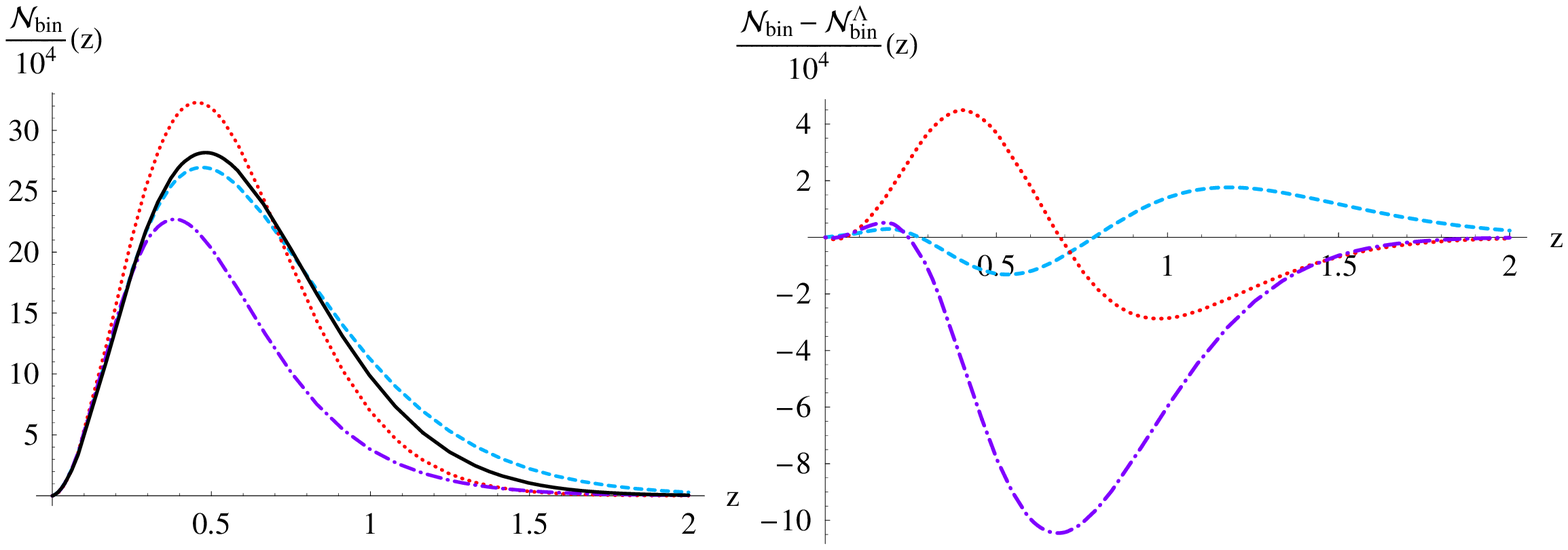,scale=0.65,angle=0}}
   \caption{\sf Evolution of number counts in mass bins with redshift
   for objects with mass within the range
   $10^{14}<M/(h^{-1}M_\odot )<10^{15}$. Notice the normalization
   factor of $\frac{\cal N}{10^4}$. Lines are the same as in figure \ref{f.omegade2}.}
   \label{f.nbin14-15}
\end{figure}
\begin{figure}[h]
   \centerline{\psfig{file=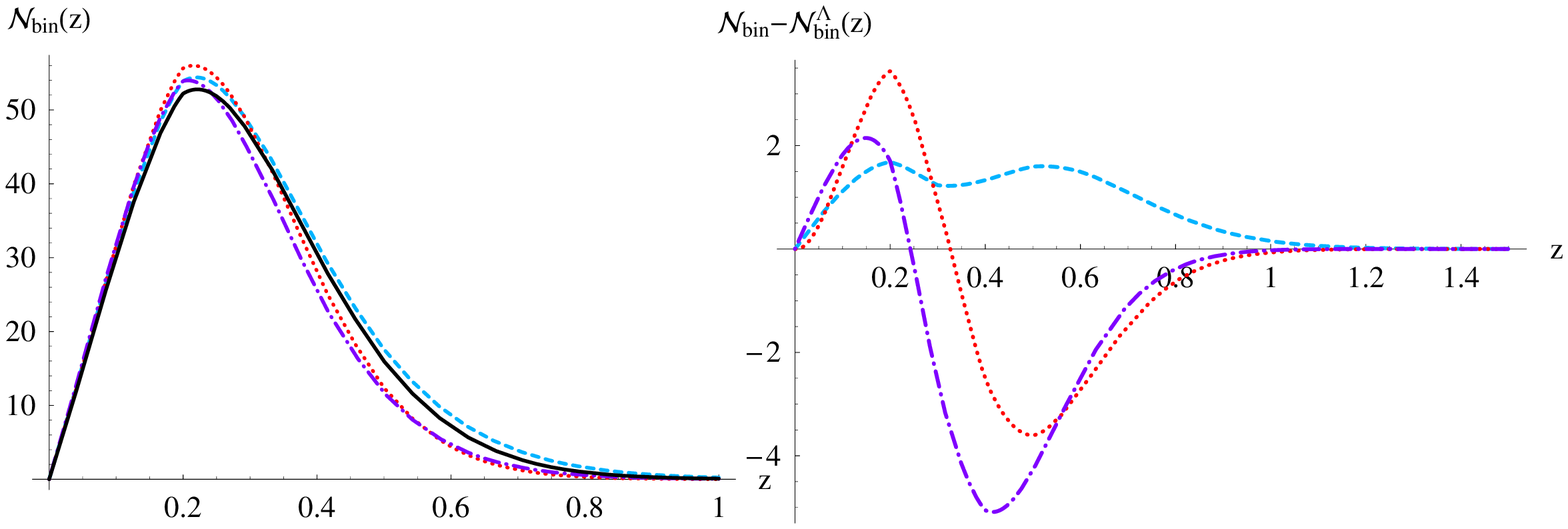,scale=0.65,angle=0}}
   \caption{\sf Same as figure \ref{f.nbin14-15} for objects with mass within
   the range $10^{15}<M/(h^{-1}M_\odot )<10^{16}$. Note that here
   ${\cal N}$ and $z$ range different from
   figure \ref{f.nbin14-15}. Lines are the same as in figure \ref{f.omegade2}.}
   \label{f.nbin15-16}
\end{figure}

The second quantity we compute is the all sky integrated number
counts above a given mass threshold, $M_{\rm inf}$, and up to
redshift $z$ \cite{nunes05}: \be
   N(z,M>M_{\rm inf})= \int_{4\pi}~d\Omega\int_{M_{\rm
   inf}}^{\infty}\int_0^z \frac{dn}{dM}~\frac{dV}{dz'd\Omega}~dM
   dz'\,.
   \label{nint}
\ee

Our knowledge of both these quantities for galaxy clusters will
improve enormously with upcoming cluster surveys operating at
different wavebands \cite{SPT}
.

The modifications caused by a dark energy component on the number
of dark matter halos are tested and confronted with a cosmological
constant $\Lambda$CDM model. We examine the effects of the
different equations of state on the number of dark matter halos in
mass bins [$M_{\rm inf}, M_{\rm sup}$] illustrating different
classes of cosmological structures, namely [$10^{13},10^{14}$],
[$10^{14},10^{15}$] and [$10^{15},10^{16}$] in units of
$h^{-1}M_\odot$.

 The number counts in mass bins, ${\cal N}^{\rm
bin}=dN/dz$, obtained from (\ref{nbin}), are shown in figures
\ref{f.nbin13-14}, \ref{f.nbin14-15} and \ref{f.nbin15-16}. In
each of these figures we plot in the left panel the actual number
counts and in the right panel we show the difference with the
fiducial $\Lambda$CDM model. Notice that the more massive
structures are less abundant and form at later times, as it should
be in the hierarchical model of structure formation. Also there is
a slight difference of the peak redshift for structure formation
in the different dark energy models considered.
\begin{figure}[h]
   \centerline{\psfig{file=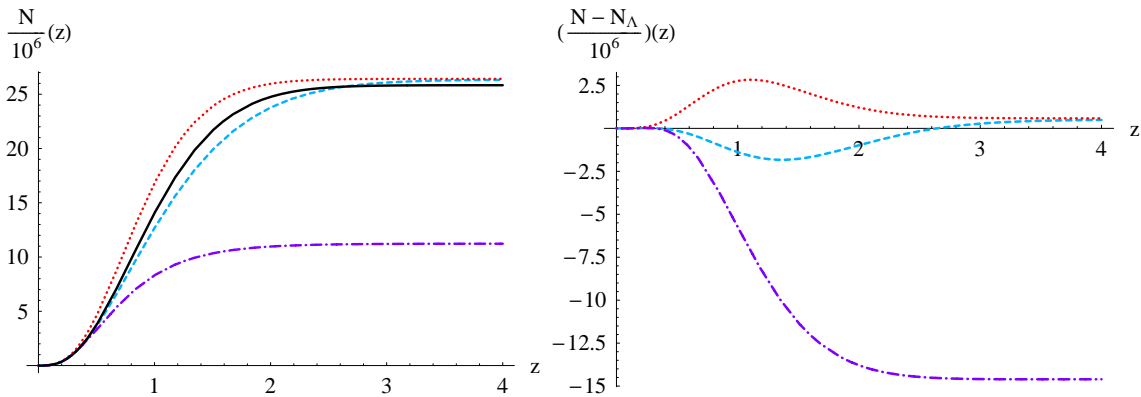,scale=0.65,angle=0}}
   \caption{\sf Evolution of the integrated number counts up to redshift $z$ for
   objects with mass above $10^{13} h^{-1}M_\odot$. Lines are the same as in figure \ref{f.omegade2}.}
   \label{f.nc13-18}
\end{figure}

 The difference among the models results from a
competition between the different volume elements and the
different growth functions. At reshifts below one the comoving
volume element has the most important role in the integral of
Eq.~(\ref{nbin}). Above this redshift the comoving volume element
does not vary significantly and the growth function becomes the
dominant source for the number counts.
\begin{figure}[t]
   \centerline{\psfig{file=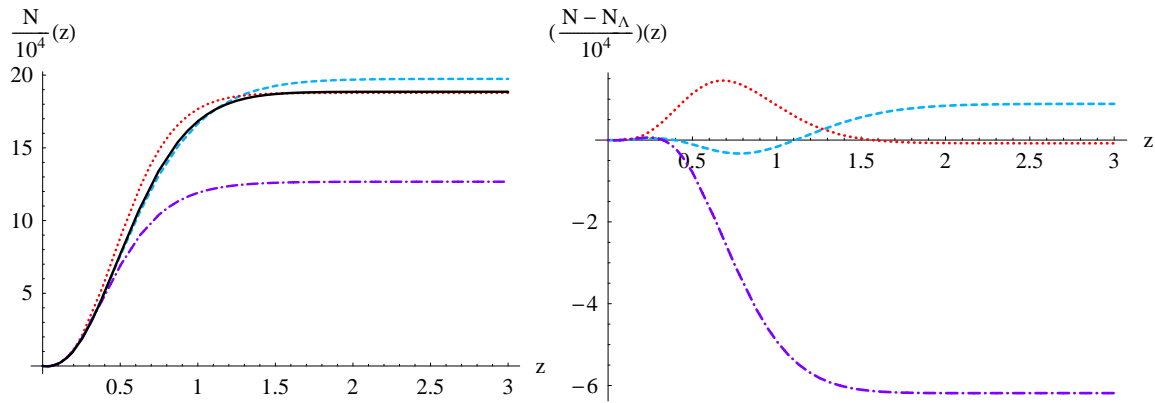,scale=0.65,angle=0}}
   \caption{\sf Evolution of the integrated number counts up to redshift $z$ for
   objects with mass above $10^{14} h^{-1}M_\odot$.Lines are the same as in figure \ref{f.omegade2}.}
   \label{f.nc14-18}
\end{figure}

\begin{figure}[t]
   \centerline{\psfig{file=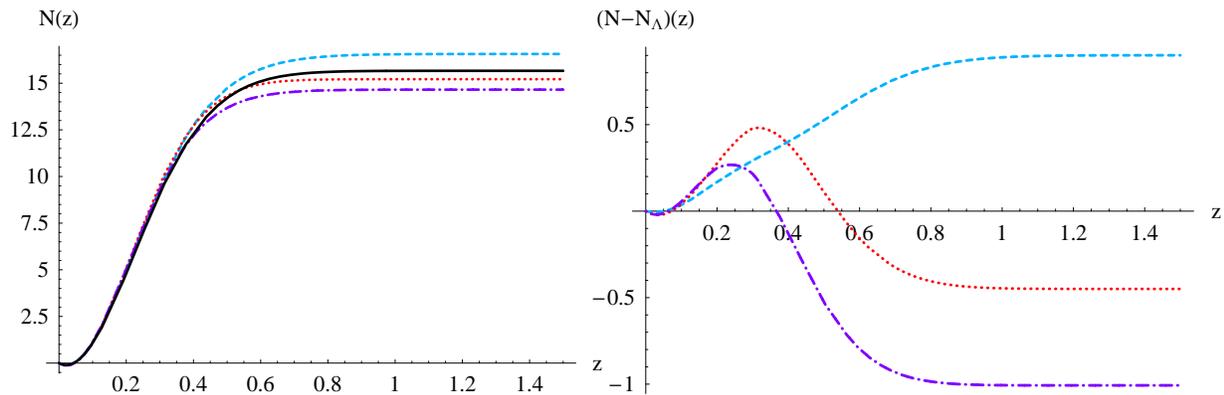,scale=0.65,angle=0}}
   \caption{\sf Evolution of the integrated number counts up to redshift $z$ for
   objects with mass above $10^{15} h^{-1}M_\odot$. Lines are the same as in figure \ref{f.omegade2}.}
   \label{f.nc15-18}
\end{figure}

An important observable quantity is the integrated number of
collapsed structures above a given mass, equation (\ref{nint}). We
present results for the integrated number counts of structures
with masses above $10^{13}h^{-1}M_\odot$, $10^{14}h^{-1}M_\odot$,
and $10^{15}h^{-1}M_\odot$. These are displayed in the figures
\ref{f.nc13-18}, \ref{f.nc14-18} and \ref{f.nc15-18}, together
with the difference with respect to the fiducial $\Lambda$CDM
model. We cut-off the integration in equation (\ref{nint}) at
$M_{\rm sup} = 10^{18} h^{-1}M_\odot$. Notice that the integrated
number has a plateau that reflects the epoch of structure
formation for a given mass. In other words, there is no formation
of structures with mass above $10^{13}h^{-1}M_\odot$,
$10^{14}h^{-1}M_\odot$, and $10^{15}h^{-1}M_\odot$ for redshifts
roughly above $z=2$, $1$ and $0.6$, respectively.

We notice that observations with accuracy of the order of $10 \%$,
either in binned or the integrated number counts, will able to
distinguish among the different models, providing important
information on the nature of dark energy.

\section{Conclusions}\label{s.conc}
$ $

We know that universe is currently accelerating and several
different mechanisms have been proposed to explain the data. In
the absence of a definitive model, an approach widely used in the
literature is to assume a given parameterization of the dark
energy equation of state. SNeIa data provides constraints in these
parameterizations but are sensitive to redshifts up to ${\cal
O}(1)$.

On the other hand, dark energy also influences the way in which
the large scale cosmological structures form. In particular,
number counts of collapsed structures is an important tool to
probe dark energy models. Since structure formation occurs at
redshifts higher than those probed by SNeIa, number counts can
provide useful complementary information in order to constraint
the parameters of the different proposed parameterizations of the
dark energy equation of state.

In this paper we showed the impact of the different
parameterizations and values for the parameters on several factors
affecting large scale structure formation. We chose to exemplify
our analysis by comparing a standard $\Lambda$CDM model with 2
parameterizations, one of them with two set of parameters
currently allowed by data.

We use the spherical collapse model in conjunction with the
Press-Schechter mass function to investigate the effect of dark
energy on the linear and non-linear growth of density
perturbations and on the number counts of collapsed structures in
different mass and redshift ranges. The dominant effect on number
counts seem to arise from the mass function, which is more
important than the comoving volume factor. The corrections arising
from the merging of clusters were shown to be small \cite{wang98}
and hence were not considered in this work.

Our results show that number counts can be useful in constraining
dark energy models. Besides showing that the unintegrated and
integrated number counts by themselves are powerful measurements,
the redshift of maximum structure formation could also be used to
differentiate the models. Observations with accuracy of the order
of $10 \%$ can be used to distinguish among the different models,
providing important information on the nature of dark energy.

On the observational side, there are new experiments planned to
start taking data on the abundance of clusters in the near future.
In particular, the South Pole Telescope survey, based  on the
detection of the Sunyaev-Zeldovich effect arising from inverse
Compton scattering of background photons off the hot intra-cluster
gas, is expected to find a large number of clusters and will be
able to determine the number counts to a high accuracy \cite{SPT}.
Hopefully these future observations will be able to discriminate
among different parameterizations of dark energy currently
proposed.

\section*{Acknowledgements}
We thank Urbano Fran\c{c}a, Reuven Opher and Ioav Waga for a careful reading of the manuscript and for providing useful suggestions and criticisms.
The work of L.L. was supported by CAPES and the work of R.R. was
partially supported by CNPq.
%

\end{document}